\documentclass{article}

\usepackage{arxiv}

\usepackage[utf8]{inputenc}
\usepackage[T1]{fontenc}
\usepackage{times}
\usepackage{graphicx}
\usepackage{amsmath}
\usepackage{amsfonts}
\usepackage{amssymb}
\usepackage{booktabs}
\usepackage{natbib}
\usepackage{doi}
\usepackage{url}
\usepackage{hyperref}

\title{Seismic full waveform inversion via a physics-guided Fourier representation neural network}

\author{
Gui Chen \thanks{Corresponding author. E-mail: chenguicup@163.com} \\
State Key Laboratory of Petroleum Resources and Engineering, \\
China University of Petroleum-Beijing, Beijing, 102249, China \\
\And
Yang Liu\\
State Key Laboratory of Petroleum Resources and Engineering,\\
China University of Petroleum-Beijing, Beijing, 102249, China
\And
Haoran Zhang\\
School of Petroleum, China University of Petroleum-Beijing at Karamay,
Karamay, 834000, China 
\And
Mi Zhang\\
School of Petroleum, China University of Petroleum-Beijing at Karamay,
Karamay, 834000, China
\And
\textsuperscript{*}Corresponding author: \href{https://orcid.org/0000-0002-6617-7365}{\raisebox{-0.15ex}{\includegraphics[height=0.9em]{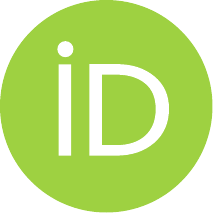}}} Gui Chen (\texttt{chenguicup@163.com})
}

\begin{document}

\maketitle

\begin{abstract}
Accurate subsurface velocity models are essential for seismic imaging, yet conventional full waveform inversion (FWI) often suffers from cycle skipping, noise sensitivity, and reliance on good initial models. We develop a physics-guided Fourier representation neural network (PGFRNN) for unsupervised acoustic FWI and simultaneous-source FWI (SSFWI), which embeds Fourier-transformed seismic data into a latent space and iteratively updates the velocity model using a softplus-approximated log-cosh (SALC) loss and a physics-guided optimizer. Numerical tests on the Overthrust model demonstrate that PGFRNN outperforms conventional L2- and SALC-loss-based FWI methods, achieving higher inversion accuracy and robustness to noise and challenging initial models.
\end{abstract}

\keywords{Full waveform inversion (FWI) \and Deep learning \and Complex-valued neural network \and Fourier transform \and Simultaneous-source FWI}

\section{Introduction}
Full waveform inversion (FWI) is a state-of-the-art technique that exploits the complete seismic wavefield to iteratively estimate accurate subsurface velocity models by minimizing the misfit between recorded and simulated data. Unlike traveltime tomography, which typically recovers smooth long-wavelength structures, FWI can resolve finer-scale velocity features.

As an ill-posed nonlinear inverse problem, FWI has motivated the development of various strategies to enhance its robustness and resolution \citep{virieux2009overview, yang2025high}. First, it is necessary to define a loss function to determine the misfit between observed and simulated data in order to calculate the velocity update gradient. A range of loss functions has been used for FWI to enhance convergence and avoid the cycle-skipping problem, including L2-norm \citep{tarantola1984inversion}, L1-norm \citep{brossier2009robust}, cross-correlation \citep{liu2017robust, zhang2019normalized}, phase-based \citep{luo2018time, bednar2007comparison}, and Wasserstein distance \citep{yang2018application, zhang2023back} objectives, each exhibiting different sensitivities to noise and amplitude variations. To further stabilize the inversion and improve resolution, regularization schemes such as Tikhonov smoothing \citep{asnaashari2013regularized, gholami2025optimal}, total variation \citep{lin2014acoustic, wo2025near}, and sparsity-promoting \citep{li2012fast, fu2023sparse} constraints are often incorporated. The choice of optimization method is closely linked to the selected loss function and regularization, with common approaches including gradient-based methods \citep{plessix2006review, xie2024full}, conjugate gradient \citep{hu2011preconditioned, pan2017accelerating}, and limited memory Broyden-Fletcher–Goldfarb–Shanno (L-BFGS) \citep{fabien2017stochastic, ye2025regularized} algorithms, each offering distinct trade-offs between computational efficiency and convergence behavior. While single-source FWI provides high-accuracy velocity estimation by updating the model for each source individually, it becomes computationally prohibitive for large-scale 3D surveys due to the need for separate forward and adjoint simulations. Simultaneous-source FWI (SSFWI) \citep{krebs2009fast, zhang2020crosstalk} addresses this limitation by stacking multiple simultaneous sources during forward modeling, thereby reducing computational cost. However, SSFWI introduces source interference (crosstalk noise), which might degrade inversion accuracy if not properly removed. Various approaches, including source encoding/decoding \citep{ben2011efficient, choi2011source, espindola2025importance}, have been developed to suppress crosstalk artifacts while maintaining high-resolution velocity estimation via SSFWI.

In recent years, deep learning (DL), a data-driven technique, has emerged as a key tool in seismology \citep{mousavi2022deep}, driving significant progress in fields such as denoising \citep{yu2019deep, dong2020denoising, chen2022dropout, zhang2025unsupervised, chen2025retrieving}, data reconstruction \citep{chen2024unsupervised, chen2025unsupervised}, resolution enhancement \citep{zhang2023improving, cheng2025self}, and seismic inversion \citep{sun2023learning, alfarhan2025robust}. DL based on neural networks (NNs) can either assist traditional FWI by predicting velocity updates or approximate gradients, or directly learn the mapping from seismic data to velocity models. From the perspective of training, DL-based FWI can be supervised, unsupervised, or self-supervised, depending on the availability of labeled velocity models. Supervised FWI requires a large and representative training dataset consisting of seismic data paired with corresponding true velocity models. \cite{yang2019deep} design a 2D convolutional NN to learn the nonlinear mapping between velocity models and seismic shot gathers. \cite{zhang2020data} develop a generative adversarial network (GAN) to learn a mapping from seismic waveforms to velocity models, in which the network simultaneously learns a data-driven regularization and imposes it on the generator to enhance inversion accuracy. Rather than depending on extensive training data \citep{sun2020ml, deng2022openfwi}, one can train a smaller NN guided by physical laws, resulting in a framework known as a physics-informed NN (PINN) \citep{raissi2019physics}, which efficiently incorporates physics equations into the learning process and has been applied successfully to FWI \citep{xu2019physics, rasht2022physics}. However, supervised DL–based FWI, even when incorporating physics-informed constraints, may still be limited by potential distribution gaps between the training data and field data.

Unsupervised DL-based FWI, which does not require paired seismic data and true velocity models for training NNs, offers a more flexible alternative. The reparameterized FWI framework \citep{wu2019parametric, zhu2022integrating, zhao2025multiscale} is an unsupervised method, which is inspired by the regularization effect of the deep image prior (DIP) \citep{ulyanov2018deep, chen2022dropout}. In this framework, regularization is introduced into FWI by representing velocity models as trainable NN parameters. Specifically, the NN generates target velocity models, and its parameters are optimized based on the misfit between modeled and observed seismic data. Compared to some reparameterization methods that use random matrices as NN inputs, some studies \citep{sun2023implicit, yang2025gabor, Ruihua11104252} feed coordinate values into the NNs to approximate the velocity model to be estimated. This type of coordinate-based approach is also referred to as implicit FWI. Additionally, while some methods use NNs solely as a regularization tool for parameterizing velocity models, recent studies directly parameterize seismic shot gathers and use observed and modeled data as NN inputs to determine their misfit in a NN-learned latent space. \cite{yang2023fwigan} introduce physics-guided FWI framework, where a wasserstein generative adversarial network (WGAN) is employed to evaluate the misfit between simulated and recorded data, while a physics-based generator optimizes velocity models using feedback from the WGAN. \cite{saad2024siamesefwi} introduce a self-supervised DL framework for FWI that employs a 2D convolutional NN to project observed and simulated shot gathers into latent spaces using shared NN weights. The discrepancy between these latent representations is quantified as the misfit, which is subsequently minimized through a physics-based optimizer to update the velocity model. In parallel to these efforts on model parameterization and misfit learning, another line of research explores SSFWI with DL. By encoding multiple shots into blended sources, SSFWI can substantially reduce computational cost, but it introduces strong crosstalk noise that hampers inversion. To address this challenge, DL techniques have been developed either as preprocessing tools to suppress crosstalk in blended data  \citep{li2022deep} or as components of the SSFWI framework itself \citep{saad2025f}. These developments illustrate the growing potential of DL to enhance FWI from both model parameterization and acquisition perspectives. However, the robustness of existing DL methods remains limited when challenged by practical issues such as noise contamination, missing low-frequency components, or crosstalk interference.

In this study, we develop an unsupervised DL framework for acoustic FWI and SSFWI under physics constraints. At its core, we introduce a physics-guided Fourier representation NN (PGFRNN) that employs complex-valued convolutions and activation functions to embed Fourier-transformed observed and modeled data into a latent space. The misfit between these latent representations is quantified using a softplus-approximated log-cosh (SALC) loss and minimized via a physics-guided optimizer to iteratively update the velocity model. We demonstrate the effectiveness of PGFRNN on the Overthrust velocity model for both FWI and SSFWI, and compare its performance with conventional L2- and SALC-loss-based FWI frameworks. Numerical experiments indicate that, while the SALC loss improves robustness to noise and poor initial models over the conventional L2 loss, PGFRNN consistently achieves superior inversion accuracy, exhibiting strong resilience to noise and the ability to recover reliable velocity models even from challenging initial models.

\section{Method}

\begin{figure}
\centering
\includegraphics[width=0.9\textwidth]{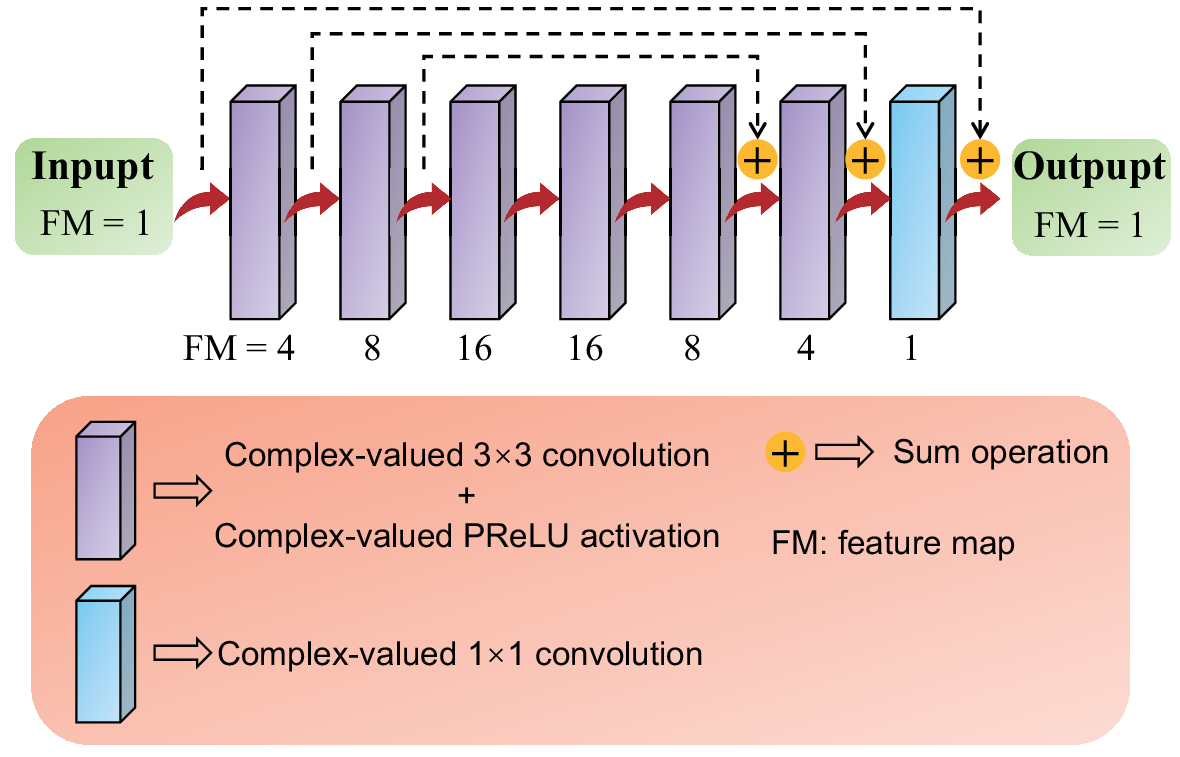}
\caption{Illustration of the designed complex-valued NN architecture.}
\label{PGFRNN}
\end{figure}

The PGFRNN framework is developed for seismic acoustic FWI and SSFWI, demonstrating robustness against poor initial models and crosstalk noise. In this framework, a complex-valued NN is designed to extract latent representations from the Fourier-transformed seismic data, both observed and modeled. As shown in Fig.~\ref{PGFRNN}, the NN takes the complex-valued spectrum obtained via the fast Fourier transform (FFT) as input and outputs the corresponding latent representation. The NN consists of six representation blocks followed by a single output block. The numbers of output feature maps (FMs) for these seven blocks are 4, 8, 16, 16, 8, 4, and 1, respectively. Each representation block contains a 2D complex-valued convolution with a kernel size of 3 and a parametric rectified linear unit (PReLU) activation function. The output block employs a 2D complex-valued convolution with a kernel size of 1 to produce a single FM, which serves as the final latent representation. Further details on the implementation of the complex-valued convolution can be found in \citep{chen2024fudlinter}. Furthermore, skip connections are incorporated between the blocks to preserve critical frequency information during the initial forward propagation. This allows the PGFRNN framework to operate like traditional FWI in the initial iterations while progressively optimizing the NN parameters in later ones, preventing the velocity model from updating in the wrong direction due to random initialization of the NN.

\begin{figure}
\centering
\includegraphics[width=0.9\textwidth]{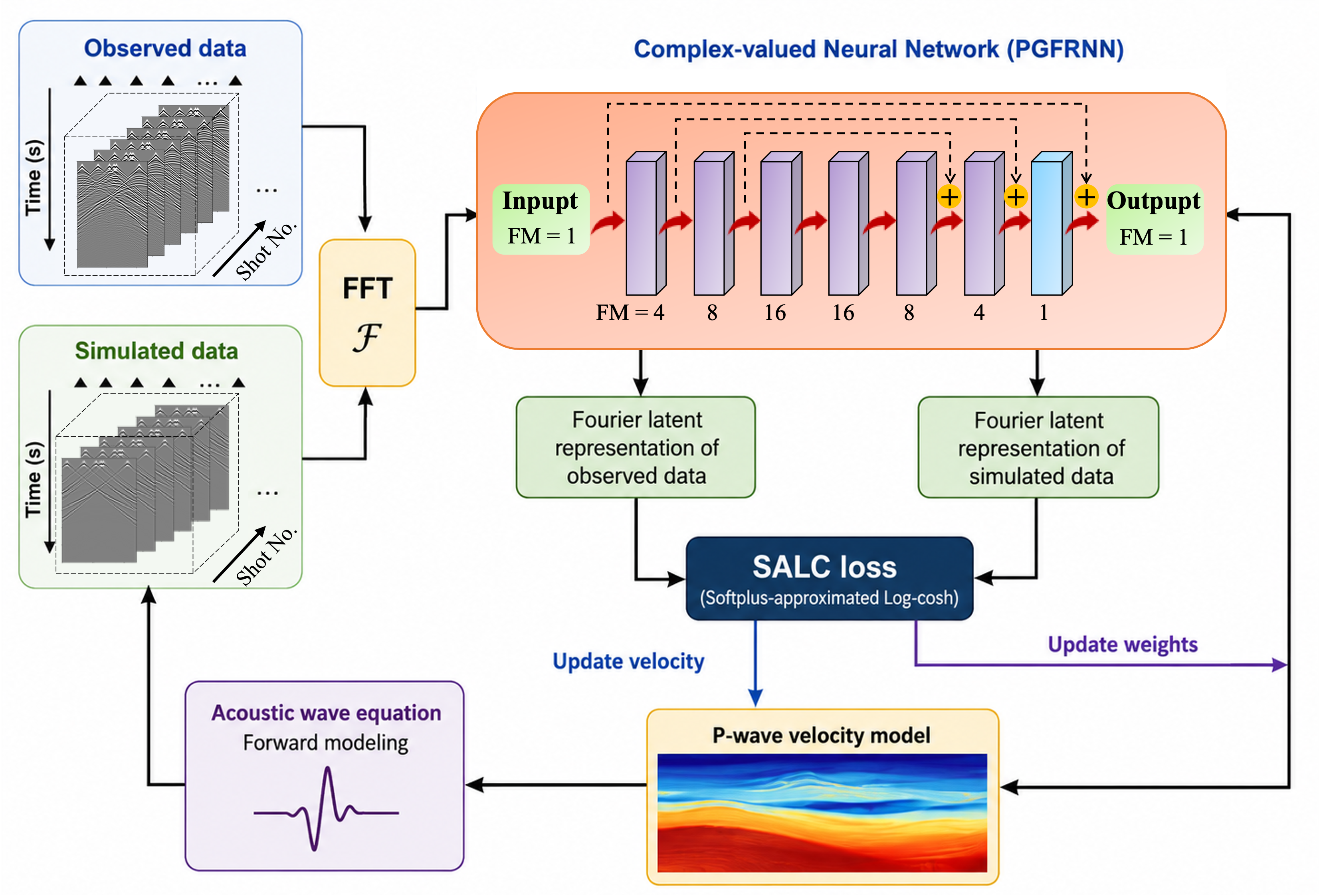}
\caption{Workflow of the proposed PGFRNN framework for acoustic FWI and SSFWI. The observed and simulated seismic data are first transformed into the frequency domain using the Fourier transform and then fed into a shared complex-valued neural network to obtain the Fourier latent representations. The misfit between the latent representations of the observed and simulated data is measured by the SALC loss, whose gradients are used to jointly update the P-wave velocity model and the network weights. The updated velocity model is subsequently used for acoustic forward modeling, forming an iterative physics-guided inversion loop.}
\label{workflow}
\end{figure}

We implement the PGFRNN inversion framework based on the designed complex-valued neural network. Fig.~\ref{workflow} illustrates the PGFRNN workflow for acoustic FWI and SSFWI. First, simulated data from the initial P-wave velocity model and observed data are transformed into the frequency domain using a 1D FFT applied to each trace. The resulting complex-valued spectra are then fed into the complex-valued NN ${f_\Theta }\left( \cdot \right)$ with trainable parameters $\Theta$ to produce complex-valued latent representations, formalized as
\begin{equation}
	{\bf{L}}{{\bf{R}}_o} = {f_\Theta }\left( {\rm FFT} \left( \mathbf{X}_o \right) \right),\qquad
	{\bf{L}}{{\bf{R}}_m} = {f_\Theta }\left( {\rm FFT} \left( \mathbf{X}_m \right) \right),
\end{equation}
where $\mathbf{X}_o$ and $\mathbf{X}_m$ denote the observed and modeled data in the time domain, and ${\bf{L}}{{\bf{R}}_o}$ and ${\bf{L}}{{\bf{R}}_m}$ represent their corresponding latent representations.

We then evaluate the misfit between the real and imaginary parts of the latent representations, which is used to generate gradients for updating the P-wave velocity model. The overall misfit function is defined as
\begin{equation}
	\Phi \left( {\bf{L}}{{\bf{R}}_o},{\bf{L}}{{\bf{R}}_m} \right) = {\rm SALC} \Big( \Re({\bf{L}}{{\bf{R}}_o}), \Re({\bf{L}}{{\bf{R}}_m}) \Big) + {\rm SALC} \Big( \Im({\bf{L}}{{\bf{R}}_o}), \Im({\bf{L}}{{\bf{R}}_m}) \Big),
\end{equation}
where $\Re(\cdot)$ and $\Im(\cdot)$ extract the real and imaginary parts of the latent representations, respectively. The operator ${\rm SALC}(\cdot)$ denotes the numerically stable softplus approximation of the log-cosh loss, defined as
\begin{equation}
	{\rm SALC}(x,y) = \frac{1}{N}\sum_{i=1}^N \Big[ {\rm softplus}\!\big(2(x_i-y_i)\big) - (x_i-y_i) - \ln 2 \Big],
\end{equation}
with ${\rm softplus}(2(x_i-y_i))=\ln(1+e^{2(x_i-y_i)})$, where $x_i$ and $y_i$ are the $i$-th elements of the input vectors $x$ and $y$, and $N$ is the total number of elements. Compared with conventional $L_1$ and $L_2$ norms, the SALC loss is less sensitive to large misfits and provides smooth, stable gradients. Applied separately to the real and imaginary parts of the complex-valued latent outputs, it allows the PGFRNN framework to fully exploit complex-valued features and generate robust gradients for both P-wave velocity and complex-valued NN parameter updates. Based on the misfit in Equation~(2), two Adam optimizers are employed: one for the velocity model and one for iteratively updating the network parameters $\Theta$.

The updated velocity model is next used to generate simulated data via the 2D constant-density acoustic wave equation:
\begin{equation}
\frac{\partial^2 p(x,z,t)}{\partial t^2} = v^2(x,z) \left( 
	\frac{\partial^2 p(x,z,t)}{\partial x^2} + \frac{\partial^2 p(x,z,t)}{\partial z^2} 
	\right) + r(t)\, \delta(x-x_s)\, \delta(z-z_s),
	\label{eq:AWE2D_pointsource}
\end{equation}
where $p(x,z,t)$ denotes the acoustic pressure field at spatial coordinates $(x,z)$ and time $t$, $v(x,z)$ is the spatially varying P-wave velocity, $r(t)$ is the source time function (e.g., Ricker wavelet), $(x_s, z_s)$ stands for the source location, and $\delta(\cdot)$ is the Dirac delta function, representing a point source. The acoustic wave equation is solved using a finite-difference scheme with second-order accuracy in time and eighth-order accuracy in space.

In this work, acoustic wave modeling and inversion procedures are implemented in PyTorch using the Deepwave toolbox \citep{richardson10deepwave}, which formulates FWI as a recurrent NN. Velocity gradients are computed via automatic differentiation, producing results comparable to the traditional adjoint-state method while facilitating seamless integration of DL frameworks. The Overthrust velocity model used in this study has a grid size of 400 $\times$ 94 with a uniform spacing of 0.03 km in both horizontal and vertical directions. Acoustic wavefields are simulated using a Ricker wavelet source with a peak frequency of 5 Hz. A total of 30 sources are regularly distributed along the surface at 0.39 km intervals, and 400 receivers are deployed at the surface with 0.03 km spacing under a free-surface condition. Seismic data are recorded for 6 s with a sampling interval of 0.003 s. For PGFRNN, the initial learning rates are set to 0.001 for the complex-valued NN optimizer and 20 for the Overthrust velocity optimizer, with a maximum of 1400 epochs for all inversion cases. For FWI examples, a batch size of 5 is used, splitting the 30 sources into six batches of five non-adjoint shot gathers each, whereas for SSFWI examples, a batch size of 1 is employed.

\begin{figure}
\centering
\includegraphics[width=0.9\textwidth]{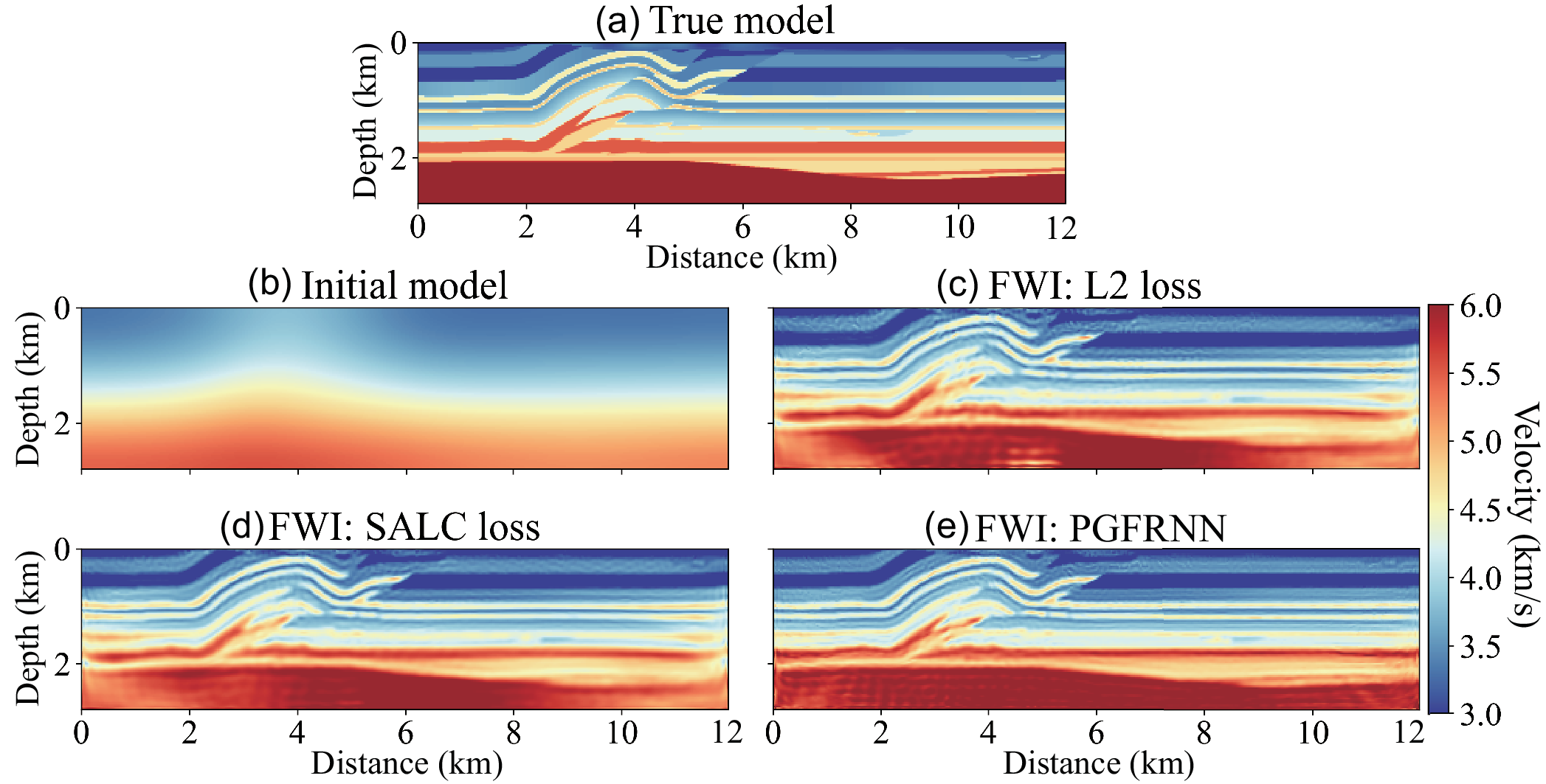}
\caption{FWI results for the Overthrust model using a Gaussian-smoothed initial model. (a) True model. (b) Gaussian-smoothed initial model. (c) Inverted model with L2 loss (TER = 23.9 dB, SSIM = 0.6420). (d) Inverted model with SALC loss (TER = 24.09 dB, SSIM = 0.6703). (e) Inverted model with the proposed PGFRNN method (TER = 28.37 dB, SSIM = 0.7246).}
\label{fwi-good}
\end{figure}

\section{Numerical experiments}
We employ two metrics to quantitatively assess the accuracy of the inverted velocity models. The first is the truth-to-error ratio (TER) in decibels (dB), defined as
\begin{equation}
\text{TER} = 10 \cdot \log_{10} \left( \frac{ \| \mathbf{v}_{\mathrm{true}} \|_2^2 }{ \| \mathbf{v}_{\mathrm{true}} - \mathbf{v}_{\mathrm{inv}} \|_2^2 } \right),
\end{equation}
where $\mathbf{v}{\mathrm{true}}$ and $\mathbf{v}{\mathrm{inv}}$ denote the true and inverted velocity models, respectively. The second metric is the structural similarity index measure (SSIM), which evaluates structural fidelity between the inverted and true velocity models:
\begin{equation}
\text{SSIM}(\mathbf{v}_{\mathrm{true}}, \mathbf{v}_{\mathrm{inv}}) = \frac{(2\mu_{\mathrm{true}} \mu_{\mathrm{inv}} + C_1)(2\sigma_{\mathrm{true,inv}} + C_2)}{(\mu_{\mathrm{true}}^2 + \mu_{\mathrm{inv}}^2 + C_1)(\sigma_{\mathrm{true}}^2 + \sigma_{\mathrm{inv}}^2 + C_2)},
\end{equation}
where $\mu_{\mathrm{true}}$ and $\mu_{\mathrm{inv}}$ denote the mean values of $\mathbf{v}_{\mathrm{true}}$ and $\mathbf{v}_{\mathrm{inv}}$, $\sigma_{\mathrm{true}}^2$ and $\sigma_{\mathrm{inv}}^2$ are their variances, and $\sigma_{\mathrm{true,inv}}$ is the covariance. Constants $C_1$ and $C_2$ are used to stabilize the division. A higher TER value indicates lower inversion error and thus better accuracy, while a higher SSIM value denotes greater structural similarity between the inverted and true velocity models, demonstrating superior preservation of structural features.

\begin{figure}
\centering
\includegraphics[width=0.9\textwidth]{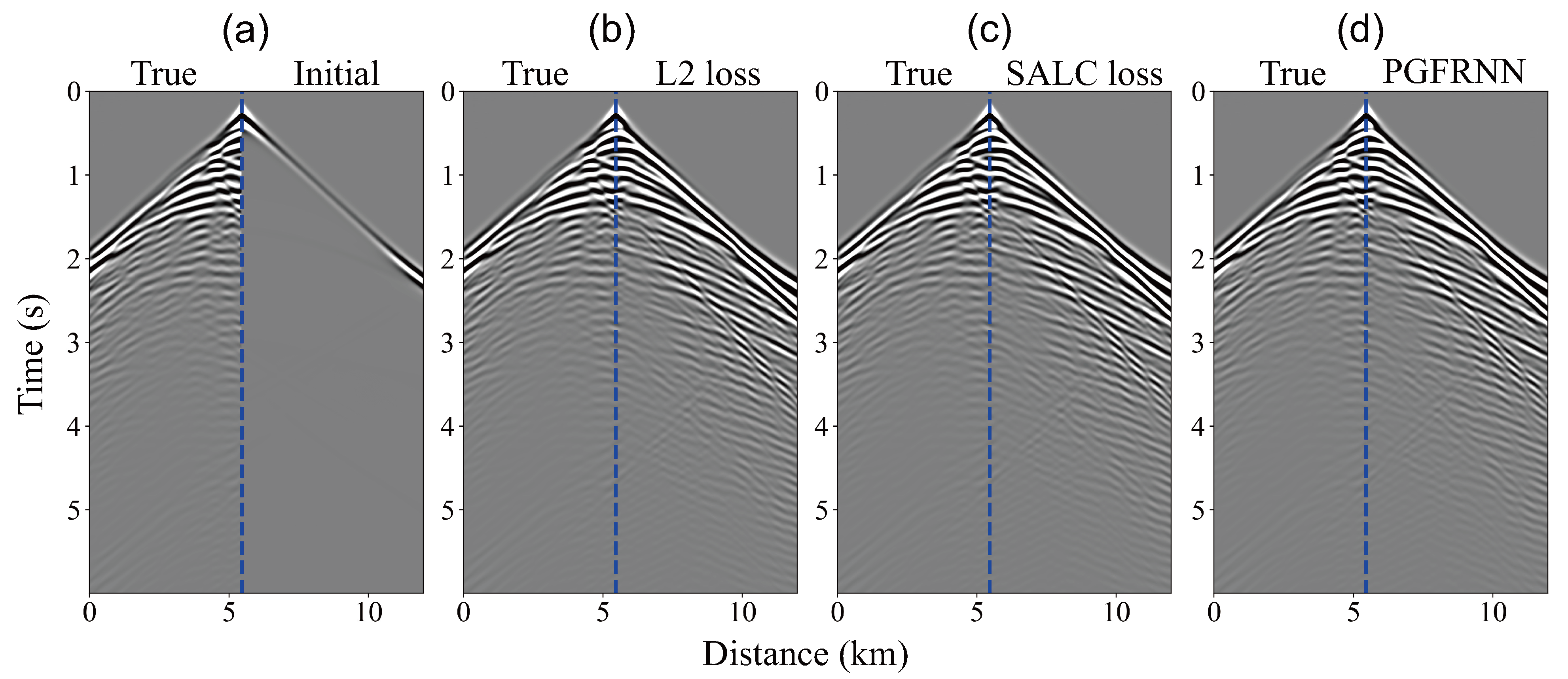}
\caption{Comparison of observed and simulated shot gathers for the Overthrust FWI example in Fig.~\ref{fwi-good}. (a) Gather corresponding to initial model. (b–d) Gathers corresponding to inverted models using L2 loss, SALC loss, and PGFRNN, respectively.}
\label{fwi-good-shot15}
\end{figure}

\begin{figure}
\centering
\includegraphics[width=0.9\textwidth]{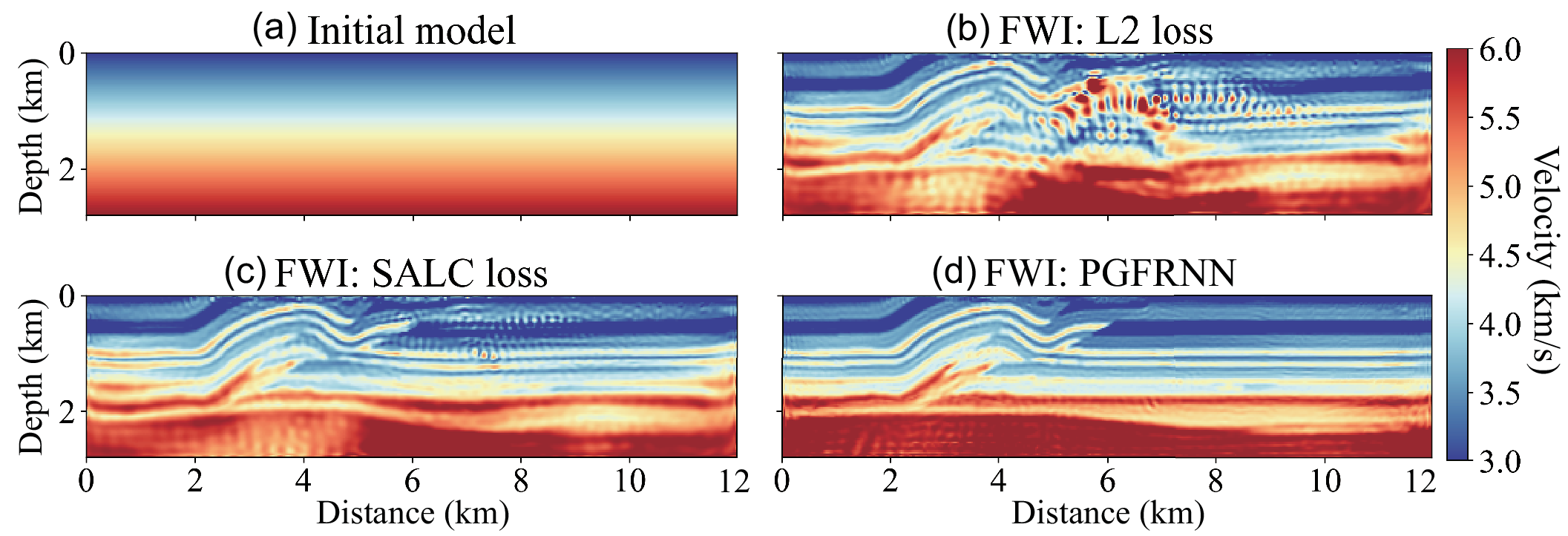}
\caption{FWI results for the Overthrust model using a linear initial model. (a) Linear initial model. (b) Inverted model with L2 loss (TER = 17.39 dB, SSIM = 0.3961). (c) Inverted model with SALC loss (TER = 21.02 dB, SSIM = 0.5001). (d) Inverted model with the proposed PGFRNN method (TER = 27.28 dB, SSIM = 0.7209).}
\label{fwi-bad}
\end{figure}

\begin{figure}
\centering
\includegraphics[width=0.9\textwidth]{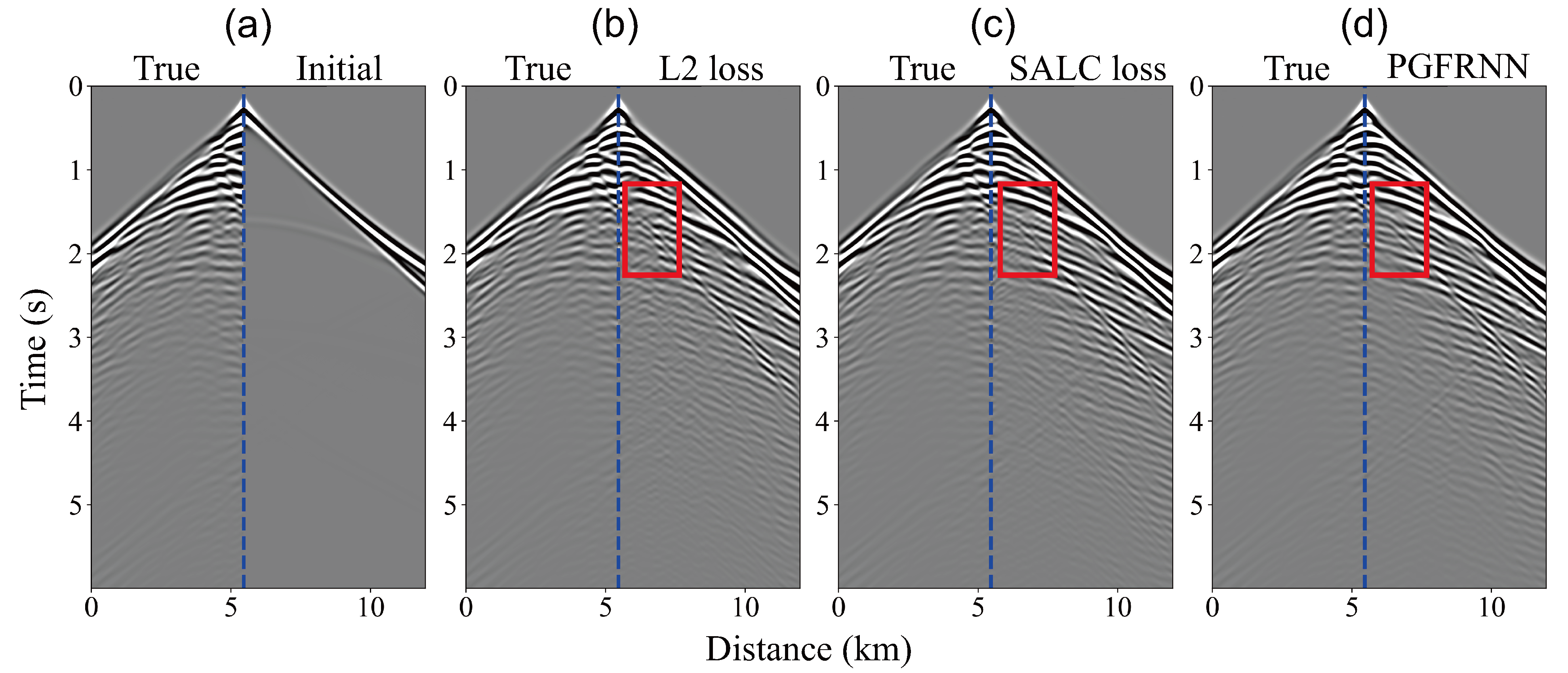}
\caption{Comparison of observed and simulated shot gathers for the Overthrust FWI example in Fig.~\ref{fwi-bad}. (a) Gather corresponding to linear initial model. (b–d) Gathers corresponding to inverted models using L2 loss, SALC loss, and PGFRNN, respectively.}
\label{fwi-bad-shot15}
\end{figure}

\begin{figure}
\centering
\includegraphics[width=0.9\textwidth]{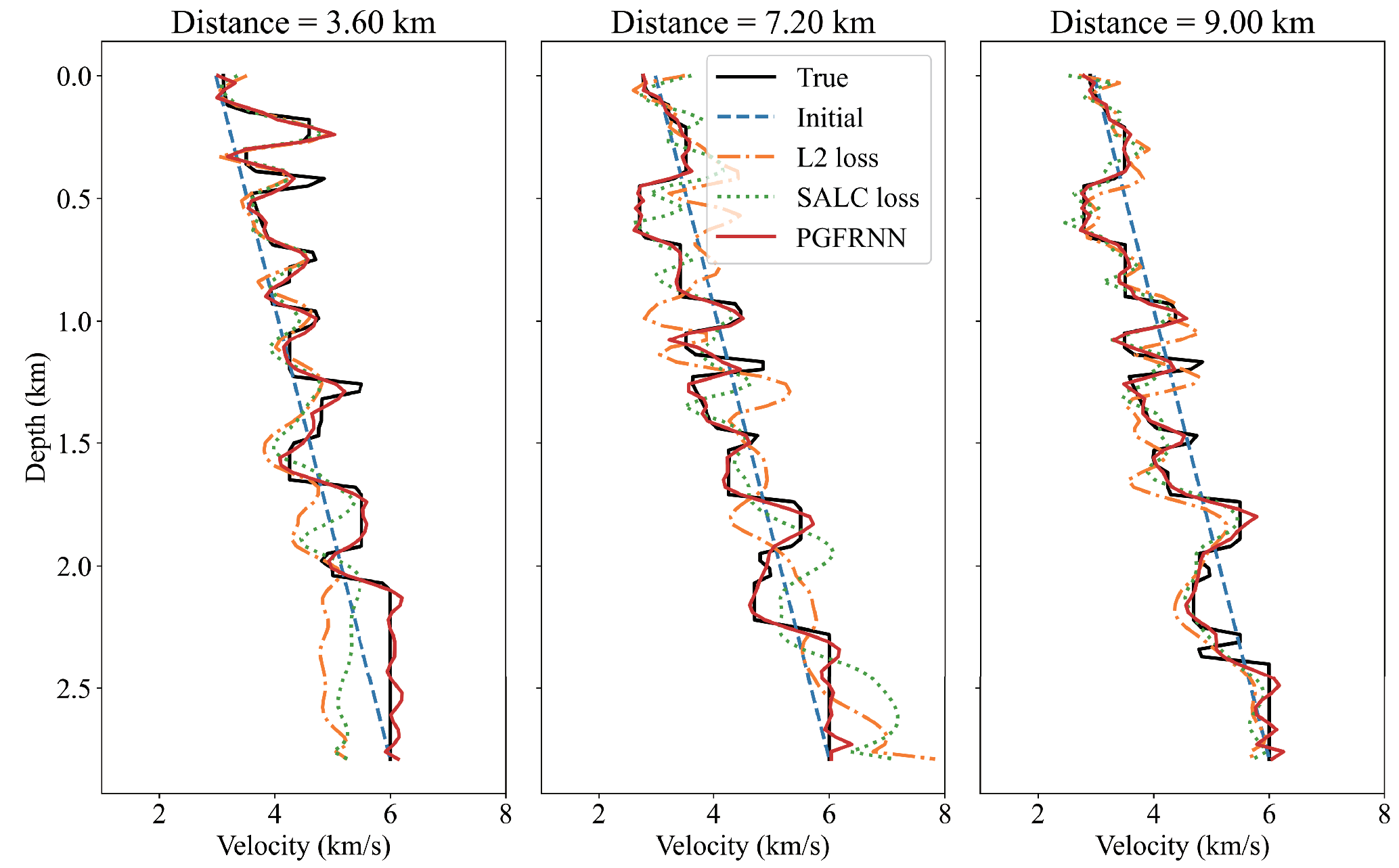}
\caption{Three velocity profiles extracted from the Overthrust FWI example in Fig.~\ref{fwi-bad}.}
\label{fwi-bad-velcurves}
\end{figure}

\begin{figure}
\centering
\includegraphics[width=0.9\textwidth]{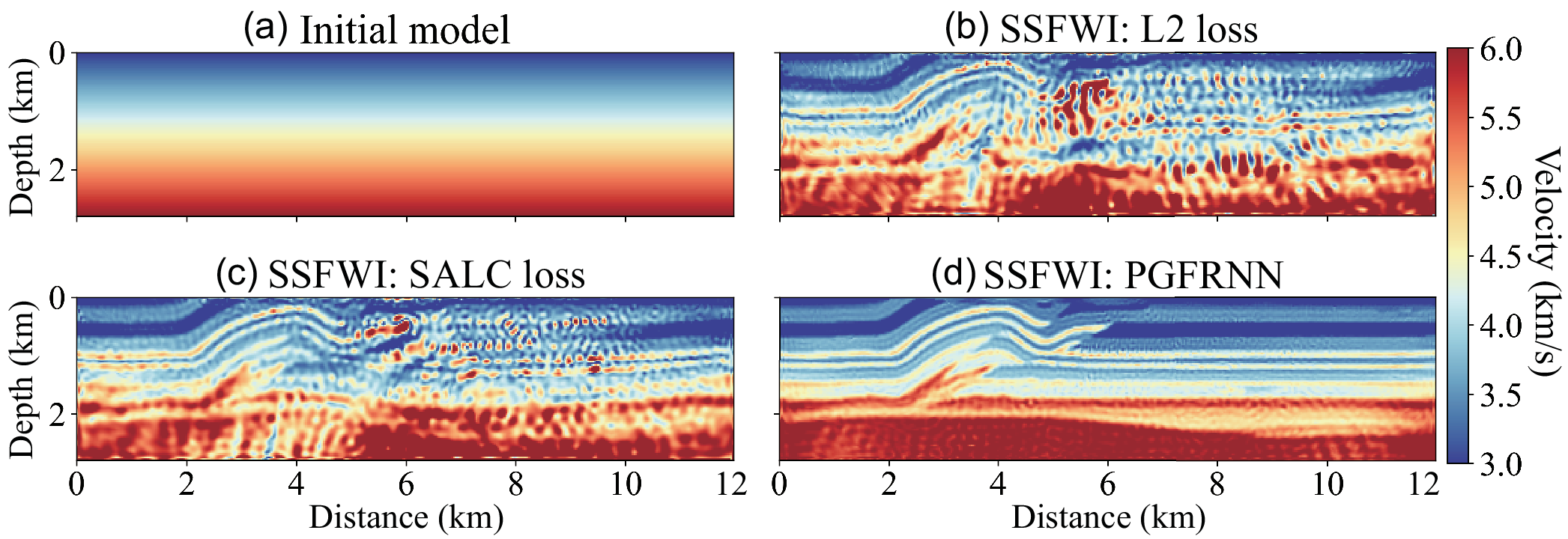}
\caption{SSFWI results for the Overthrust model using a linear initial model. (a) Linear initial model. (b) Inverted model with L2 loss (TER = 15.48 dB, SSIM = 0.2484). (c) Inverted model with SALC loss (TER = 16.75 dB, SSIM = 0.2980). (d) Inverted model with the proposed PGFRNN method (TER = 27.33 dB, SSIM = 0.7146).}
\label{ssfwi-bad}
\end{figure}

\begin{figure}
\centering
\includegraphics[width=0.8\textwidth]{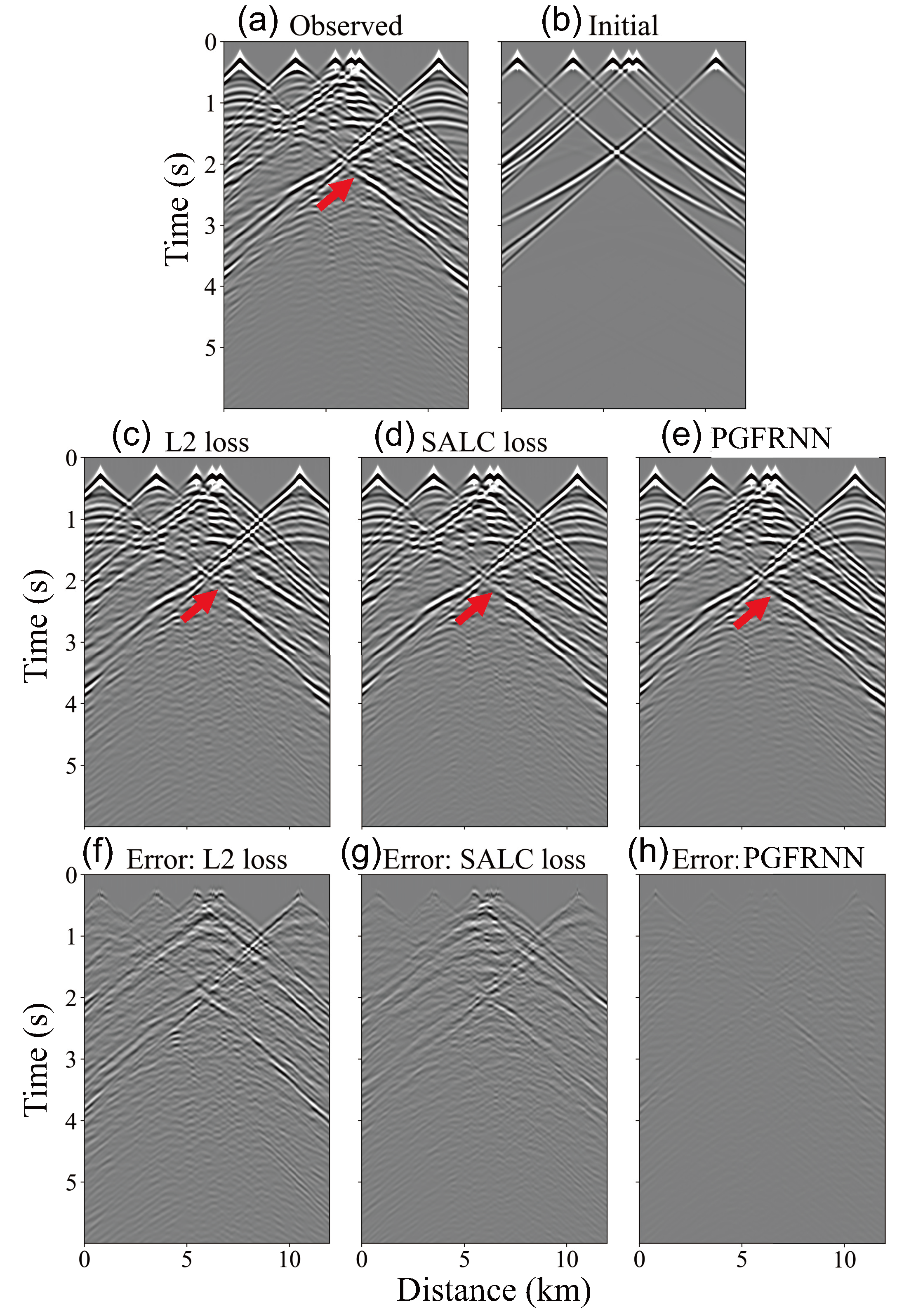}
\caption{Comparison of observed and simulated super-shot gathers for the Overthrust SSFWI example in Fig.~\ref{ssfwi-bad}. (a) Gather corresponding to true model. (b) Gather corresponding to linear initial model. (c–e) Gathers corresponding to inverted models using L2 loss, SALC loss, and PGFRNN, respectively. (f–h) Corresponding super-shot gather errors with MSE = 0.31, 0.13, and 0.005, respectively.}
\label{ssfwi-bad-shot4}
\end{figure}

\subsection{FWI}
We first apply the proposed PGFRNN method to the Overthrust model to evaluate its performance. The initial velocity model is constructed by smoothing the true model with a Gaussian filter. Figs.\ref{fwi-good}(a) and \ref{fwi-good}(b) show the true and initial models, respectively. Figs.\ref{fwi-good}(c) and \ref{fwi-good}(d) present the inversion results obtained by conventional FWI framework using the L2 and SALC losses, respectively, while Fig.\ref{fwi-good}(e) displays the inverted model from PGFRNN. All three methods are able to reconstruct the main geological structures, such as the fault systems. As shown in Fig.\ref{fwi-good-shot15}, the synthetic data generated from the inverted velocity model using the three methods show a close match to the observed seismic data in terms of both waveform and amplitude. The TER values of the inverted model using L2 and SALC losses within the conventional FWI framework are 23.90 and 24.09 dB, respectively, with corresponding SSIM values are 0.6420 and 0.6703. In contrast, the PGFRNN framework achieves a TER value of 28.37 dB and an SSIM value of 0.7246, indicating substantial improvements in both fidelity and structural similarity over the conventional L2 and SALC-based FWI approaches.

To further investigate the robustness of the proposed PGFRNN framework, we consider a poor initial model defined as a simple linearly increasing velocity with depth, as shown in Fig.\ref{fwi-bad}(a). Compared to the true model in Fig.\ref{fwi-good}(a), this linear initial model lacks any structural information, making the inversion highly challenging and prone to cycle skipping. The inverted models obtained using conventional FWI with the L2 and SALC losses are displayed in Figs.\ref{fwi-bad}(b) and \ref{fwi-bad}(c), respectively. The inversion using the conventional L2 objective appears to be trapped in a local minimum. Its recovered model shows pronounced oscillatory artifacts and misses several fault features. This behavior is consistent with cycle-skipping and with the L2 objective’s sensitivity to large misfits, which can amplify phase mismatches and produce spurious high-frequency content. By contrast, the SALC objective demonstrates greater robustness to large misfits and produces a more reliable result, successfully recovering the major fault structures with only minor irregularities near fault zones. The PGFRNN inversion result, shown in Fig.\ref{fwi-bad}(d), demonstrates improvements over the conventional L2 and SALC objectives. Even starting from the simple linear initial model, PGFRNN accurately reconstructs the major structural features, including layer boundaries and fault zones, while substantially reducing oscillatory artifacts. The inverted models using conventional FWI with the L2 and SALC objectives achieve TERs of 17.39 and 21.02 dB and SSIMs of 0.3961 and 0.5001, respectively. In comparison, PGFRNN reaches a TER value of 27.28 dB and an SSIM value of 0.7209.
For further comparison, shot gathers generated from the three inverted models are shown in Fig.\ref{fwi-bad-shot15}. The regions highlighted by red boxes indicate that PGFRNN provides superior reflection waveform matching relative to the two benchmark methods. Additionally, vertical velocity profiles extracted at three locations (Fig.\ref{fwi-bad-velcurves}) show that the PGFRNN-inverted model closely follows the true velocity structure.

\begin{figure}
\centering
\includegraphics[width=0.9\textwidth]{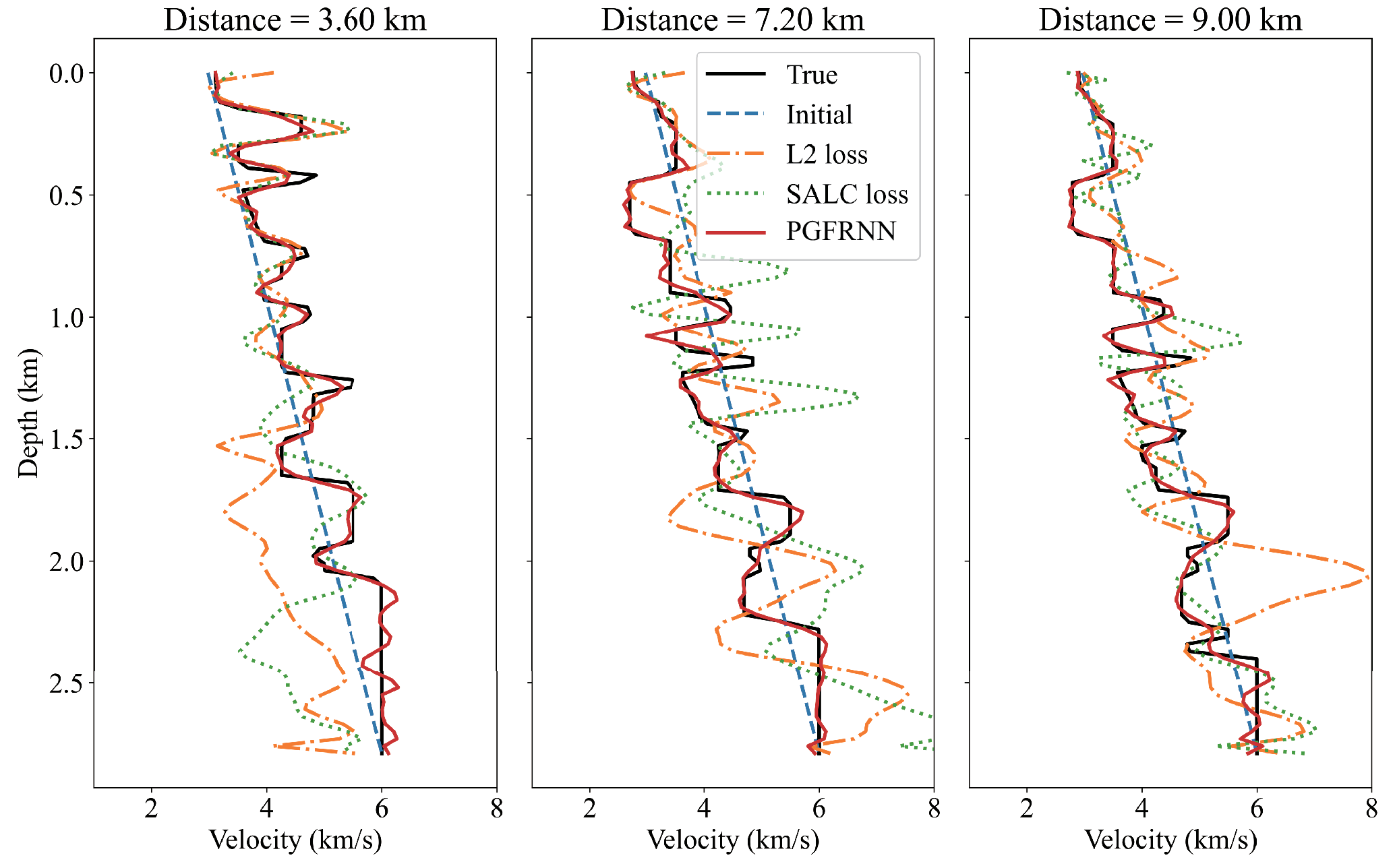}
\caption{Three velocity profiles extracted from the Overthrust SSFWI example in Fig.~\ref{ssfwi-bad}.}
\label{ssfwi-bad-velcurves}
\end{figure}

\subsection{SSFWI}
We next evaluate the performance of PGFRNN under a SSFWI setup, starting from the same simple linear initial velocity model (Fig.\ref{ssfwi-bad}(a)). In this setup, six simultaneous sources are randomly grouped into a single super-shot, producing five super-shots to cover all 30 sources. While SSFWI theoretically accelerates FWI by a factor of six, it introduces potential crosstalk noise, increasing the inversion challenge.
The inverted models obtained using conventional L2 and SALC objectives are shown in Figs.\ref{ssfwi-bad}(b) and \ref{ssfwi-bad}(c), respectively. Both methods struggle under the SSFWI setup, exhibiting cycle skipping and incomplete recovery of fault structures due to interference between overlapping sources. In contrast, PGFRNN (Fig.~\ref{ssfwi-bad}(d)) successfully reconstructs the major layer interfaces and fault features, while preserving fine-scale structures and mitigating crosstalk effects.
Quantitative evaluation shows that PGFRNN achieves a TER value of 27.33 dB and an SSIM value of 0.7146, substantially outperforming the L2- and SALC-based SSFWI results.

Fig.\ref{ssfwi-bad-shot4} shows the simulated super-shot gathers generated from the inverted models obtained using the three methods, along with their differences relative to the observed true data. As highlighted by the red arrows, the super-shot gather corresponding to the PGFRNN method exhibits higher waveform similarity to the true data compared to the L2 and SALC methods. Quantitatively, the MSE value of the gather associated with PGFRNN is only 0.005, compared to 0.31 and 0.13 for the L2- and SALC-based gathers, respectively, highlighting the superior data-fitting performance of PGFRNN under the SSFWI setup.
Moreover, three vertical velocity profiles extracted from the inverted models at distances of 3.6 km, 7.2 km, and 9 km are shown in Fig.\ref{ssfwi-bad-velcurves}. The profiles indicate that PGFRNN closely matches the true velocity model, accurately recovering both major interfaces and fine-scale structures, whereas the L2 and SALC inversions deviate noticeably, exhibiting strong oscillations.

\begin{figure}
\centering
\includegraphics[width=0.9\textwidth]{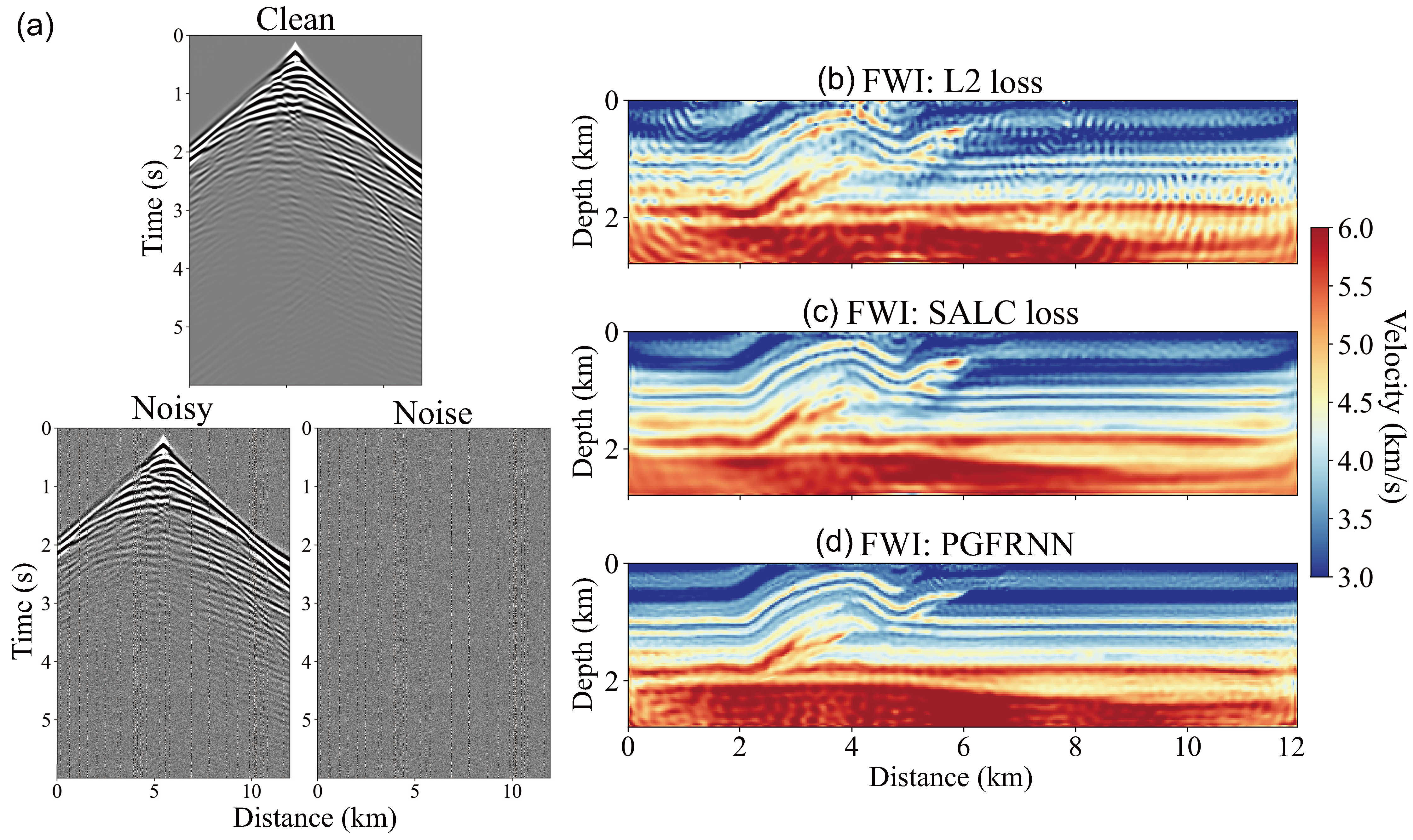}
\caption{FWI results for the Overthrust example in Fig.~\ref{fwi-good} with noisy observed data contaminated by random and erratic noise. (a) Clean observed gather, noisy observed gather, and added noise. (b) Inverted model with L2 loss (TER = 20.04 dB, SSIM = 0.3636). (c) Inverted model with SALC loss (TER = 21.05 dB, SSIM = 0.4720). (d) Inverted model with PGFRNN (TER = 24.73 dB, SSIM = 0.6600).}
\label{fwi-good-noisy}
\end{figure}

\begin{figure}
\centering
\includegraphics[width=0.9\textwidth]{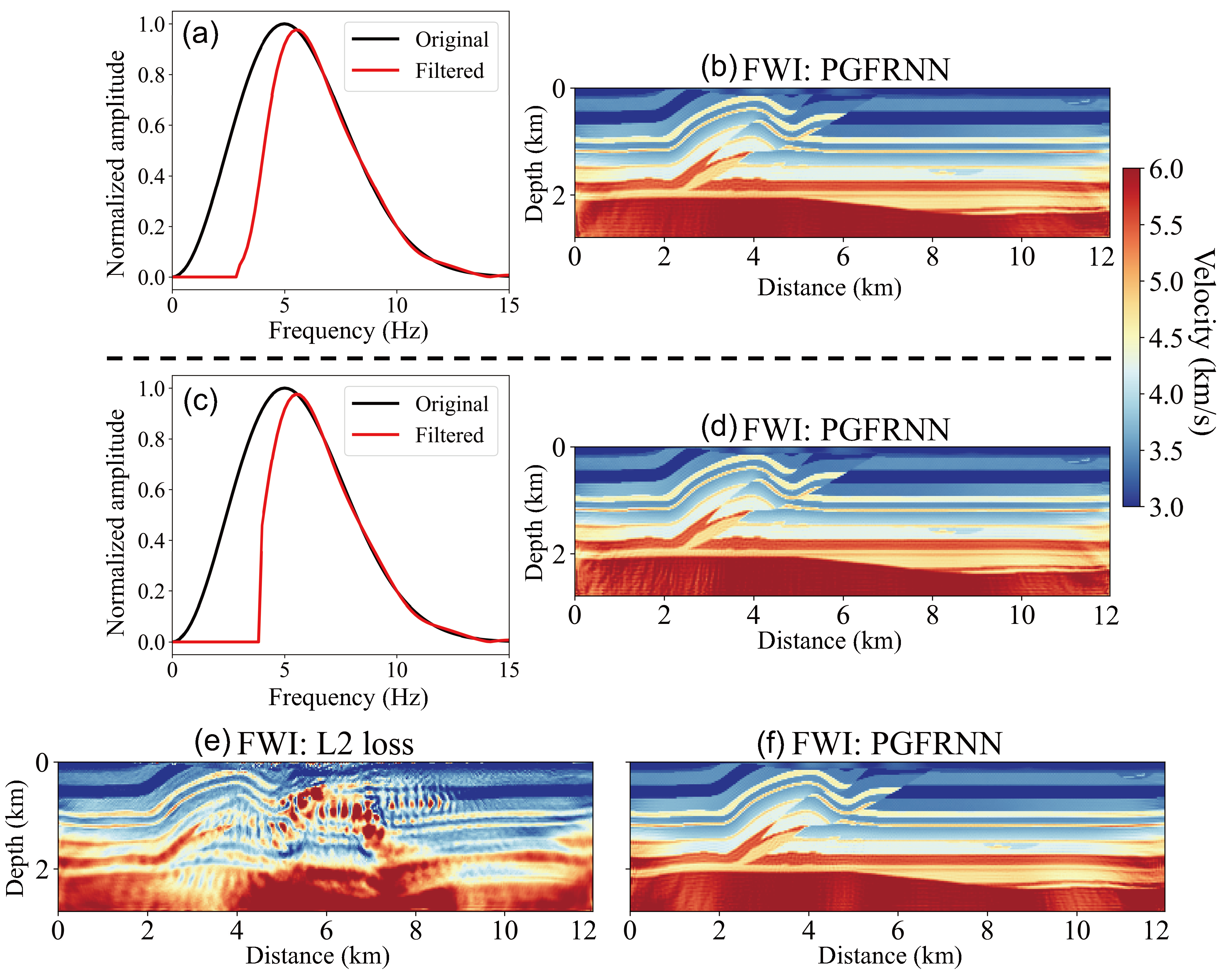}
\caption{FWI results for the Overthrust model with high-pass filtered Ricker wavelets.
(a) Amplitude spectrum of a 3 Hz high-pass filtered wavelet.
(b) Inverted model obtained by PGFRNN (TER = 30.56 dB, SSIM = 0.8606) using the 3 Hz high-pass filtered wavelet and the Gaussian-smoothed initial model in Fig.\ref{fwi-good}(b).
(c) Amplitude spectrum of a 4 Hz high-pass filtered wavelet.
(d) Inverted model obtained by PGFRNN (TER = 30.09 dB, SSIM = 0.8486) using the 4 Hz high-pass filtered wavelet and the Gaussian-smoothed initial model in Fig.\ref{fwi-good}(b).
(e) Inverted model obtained by L2 loss (TER = 16.15 dB, SSIM = 0.3973) using the 4 Hz high-pass filtered wavelet and the linear initial model in Fig.\ref{fwi-bad}(a).
(f) Inverted model obtained by PGFRNN (TER = 28.19 dB, SSIM = 0.8309) using the 4 Hz high-pass filtered wavelet and the linear initial model in Fig.\ref{fwi-bad}(a).}
\label{fwi-good-filter}
\end{figure}

\section{Discussion}

\subsection{Impact of noise}
In practical applications, observed data are often contaminated by noise rather than being perfectly clean. To assess the robustness of the proposed PGFRNN method under noisy conditions, we add Gaussian random noise and erratic noise to the observed data for FWI. Fig.\ref{fwi-good-noisy}(a) shows a shot gather and its noisy counterpart with an SNR value of -0.61 dB, where some weak reflections are obscured by noise. The inverted models obtained using the L2 loss, SALC loss, and PGFRNN are shown in Figs.\ref{fwi-good-noisy}(b)–\ref{fwi-good-noisy}(d), with TER values of 20.04, 21.05, and 24.73 dB, and corresponding SSIM values of 0.3636, 0.4720, and 0.6600, respectively.
Compared with the noise-free case, the accuracy of all three methods decreases. The conventional L2 objective is affected by strong noise, producing a highly non-smooth result, while the SALC loss yields smoother gradients and a more stable inversion. By contrast, PGFRNN achieves the most reliable inversion, demonstrating superior robustness to noise.

\subsection{Impact of missing low-frequency components}
Low-frequency information in FWI generally helps mitigate cycle skipping and guides inversion toward the global minimum. However, in practical acquisition scenarios, low-frequency components are often missing due to limitations of seismic sources, receiver responses, and strong ambient noise.

To evaluate the robustness of PGFRNN under missing low-frequency conditions, we first perform two experiments using the Gaussian-smoothed initial velocity model with high-pass filtered Ricker wavelets. In the first case, frequencies below 3Hz are removed (Fig.\ref{fwi-good-filter}(a)), and in the second case, the cutoff is increased to 4Hz (Fig.\ref{fwi-good-filter}(c)). The corresponding PGFRNN inversions, shown in Figs.~\ref{fwi-good-filter}(b) and \ref{fwi-good-filter}(d), both exhibit improved structural recovery compared with the full-frequency inversion. This improvement arises because the smoothed initial model already provides a reasonable large-scale velocity trend, while the removal of very low-frequency components suppresses overly smooth large-scale updates, allowing mid- to high-frequency information to dominate the reconstruction of fine-scale structures in the Overthrust model.

In the third experiment, we use the 4Hz high-pass filtered Ricker wavelet combined with the linear initial velocity model, and compare PGFRNN with conventional L2-based inversion. The corresponding inverted models are shown in Figs.\ref{fwi-good-filter}(e) and \ref{fwi-good-filter}(f), respectively. The L2-loss inversion exhibits strong oscillatory artifacts and fails to recover major fault structures. In contrast, PGFRNN produces significantly smoother and more accurate results, and its inversion result with missing low-frequency components and the linear initial model is also substantially better than the full-frequency inversion using the same linear model. This demonstrates that PGFRNN can effectively exploit mid- to high-frequency information to reconstruct the main interfaces and fault structures, even with poor initial models and absent low-frequency data.

Overall, these experiments demonstrate that PGFRNN is robust to missing low-frequency components. In the Overthrust model, the removal of very low-frequency information can even improve PGFRNN-based FWI. This model-dependent improvement arises from the combination of a reasonably accurate initial model and the dominance of mid- to high-frequency components in reconstructing strong velocity contrasts and fault structures.

\section{Conclusion}
We introduce PGFRNN, an unsupervised deep learning framework for acoustic FWI and SSFWI that embeds Fourier-transformed seismic data into a latent space and iteratively updates velocity models via a SALC loss and a physics-guided optimizer. Tests on the Overthrust model demonstrate that PGFRNN outperforms conventional L2- and SALC-loss-based FWI, achieving higher accuracy, robustness to noise and missing low-frequency data, and effective mitigation of crosstalk noise in simultaneous-source setups.

\section*{Acknowledgments}
This work was supported by the National Natural Science Foundation of China (Grant No. 42274147).

\bibliographystyle{unsrtnat}
\bibliography{reference}

\end{document}